\documentclass{bioinfo}
\copyrightyear{2012}
\pubyear{2012}

\usepackage{graphicx}
\usepackage{url}
\usepackage{subfigure}
\usepackage{multirow}
\usepackage{threeparttable}
\usepackage{ifthen}
\usepackage{caption2}

\begin{document}
\firstpage{1}

\title[On unbiased performance evaluation for protein inference]{On unbiased performance evaluation for protein inference}
\author[He \textit{et~al}]{Zengyou He \footnote{to whom correspondence should be addressed}, Ting Huang and Peijun Zhu\,}
\address{School of Software, Dalian University of Technology, Dalian 116620, China.\\}

\history{Received on XXXXX; revised on XXXXX; accepted on XXXXX}

\editor{Associate Editor: XXXXXXX}
\maketitle

\textbf{Contact: }\href{zyhe@dlut.edu.cn}{zyhe@dlut.edu.cn}

\section{Background}

This letter is a response to the comments of \citet{Serang12} on
\citet{HH12} in \emph{Bioinformatics}. \citet{Serang12} claimed that
the parameters for the Fido algorithm should be specified using the
grid search method in \citet{Fido} so as to generate a deserved
accuracy in performance comparison. It seems that it is an argument
on parameter tuning. However, it is indeed the issue of how to
conduct an unbiased performance evaluation for comparing different
protein inference algorithms. In this letter, we would explain why
don't we use the grid search for parameter selection in \citet{HH12}
and show that this procedure may result in an over-estimated
performance that is unfair to competing algorithms. In fact, this
issue has also been pointed out by \citet{LR12}.

\section{Model Selection and Assessment in Protein Inference}
Machine learning is a cornerstone of modern bioinformatics.
Meanwhile, an unbiased performance evaluation is undoubtedly the
cornerstone of machine learning research and applications
(\citealp{GT10}), which provides a clear picture of the strengths
and weaknesses of existing approaches.

In the real world application of machine learning methods, there are
two closely related and separate problems: model selection and model
assessment (\citealp{book}). In model selection, we estimate the
performance of different models in order to choose the best one. In
model assessment or performance evaluation, we test the prediction
error of a final model obtained from the model selection process.

The protein inference problem is an instance of prediction task in
machine learning as well, as shown in Fig.1. In model selection, we
use the peptide-protein bipartite graph as the input to find a
``best'' inference model that produces a vector $\hat{Y}$
\citep*{HWY12}. Each element in $\hat{Y}$ can be either the
probability/score that each protein is present or the presence
status of each protein (true or false). In model assessment, we
compare the predicted vector $\hat{Y}$ with ground-truth vector $Y$
to obtain the performance estimates. This is the correct procedure
for evaluating and comparing protein inference algorithms.

In contrast, one possible mistake in an incorrect procedure is
illustrated at the top of Fig.1: the partial or whole ground-truth
vector $Y$ is used in the model selection process of the protein
inference algorithms. The problem is that the inference algorithms
have an unfair advantage since they ``have already seen'' the
absence/presence information in $Y$ that should only be available
during model assessment. In other words, the ground-truth
information has leaked to the model selection phase. As a result,
the performance estimates of inference algorithms will be over
optimistic. This phenomenon is essentially analogous to the
selection bias observed in classification or regression due to
feature selection over all samples prior to performance evaluation
(\citealp{SFK10}).

\begin{figure}
        \centering
                \includegraphics[width=0.48\textwidth]{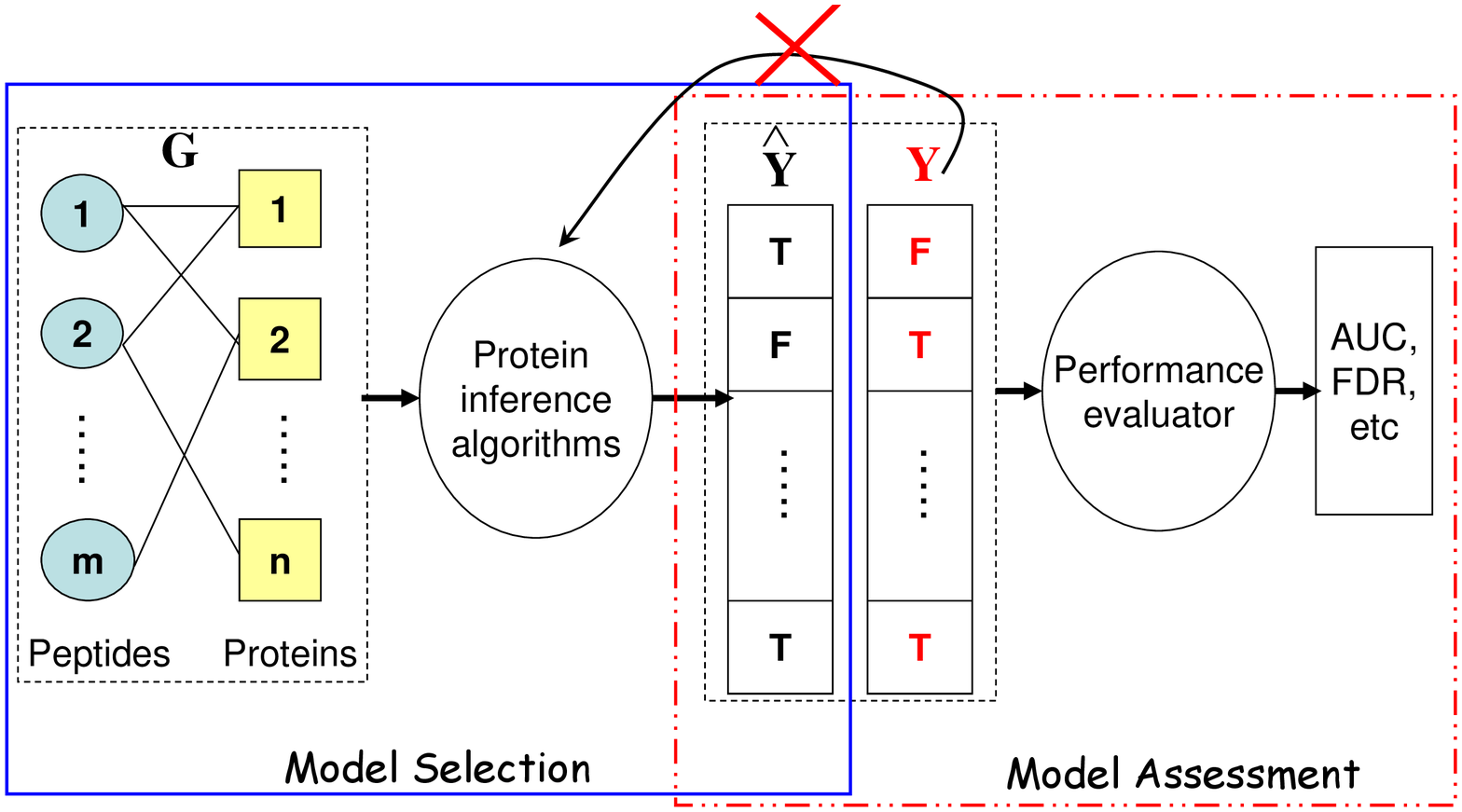}
                \caption{The correct and incorrect procedure for assessing the performance of protein inference algorithms.
In model selection, we cannot use any ground-truth information that
should be only visible in the model assessment stage. Otherwise, we
may over-estimate the actual performance of inference algorithms.}
                \label{Fig1}
\end{figure}

\begin{figure}
\centering
\includegraphics[width=0.4\textwidth]{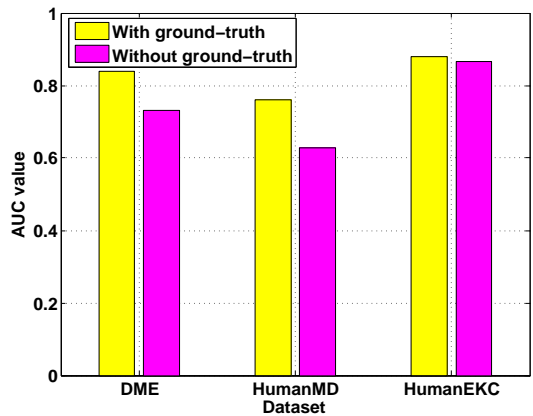}
\caption{The effect of using the ground-truth information in the
grid search procedure of Fido. The grid search procedure finds a set
of parameters automatically with the ground-truth labels of
candidate proteins as input. Note that such presence/absence
information of proteins should not be visible to the inference
algorithms if they are once again used in the performance evaluation
stage for comparing different algorithms. To mimic the situation
that the ground-truth information is unavailable, we assign a zero
weight to $ROC_{50}$ in the grid search method and calculate the
average area under curve (AUC) value as the performance index of
``without ground-truth''.} \label{Fig2}
\end{figure}

According to the description in \citet{Fido} and the source codes of
Fido, the grid search procedure chooses the set of parameters that
jointly maximizes the $ROC_{50}$ score (the average sensitivity when
allowing between zero and 50 false positives) and minimizes the mean
squared error (MSE) from an ideally calibrated probability. Clearly,
it has used the ground-truth information (true and false positive
labels)\footnote{In the target-decoy database search and evaluation
strategy, a protein is regarded as a true positive if it comes from
the target database and as a false positive otherwise. Therefore,
the set of target/decoy labels is used as the set of ground-truth
labels in this context, although some target proteins may be false
positives. } that should only be available in the model assessment
stage. In particular, the grid search procedure selects parameters
using the $ROC_{50}$ score as a key factor, which is directly
related to the final performance index in the model assessment
stage. Therefore, it is highly possible that over-fitting occurs,
i.e., the use of grid search will lead to a performance
overestimation.

To check if the grid search method will lead to an over-optimistic
performance, we conduct the following experiment. As this procedure
can control for how much weight should be given to $ROC_{50}$ and
how much weight should be given to MSE in model selection, we first
assign a zero weight to $ROC_{50}$ to roughly mimic the situation
that the ground-truth information is invisible so that no
over-estimation occurs. Then, we compare its performance with that
given by the algorithm when the default non-zero weight is used for
$ROC_{50}$. As shown in Fig.2, the performance of Fido will be
deceased when the ground-truth information (in terms of $ROC_{50}$)
is not used in model selection. One may argue that we cannot fully
attribute the performance gain in grid search to the incorrect use
of ground-truth information, but at least, it will be unfair to
other competing algorithms in performance comparison.

\section{Summary}

The fact that over-fitting at the level of model selection can have
a very substantial deleterious effect in performance evaluation has
been widely discussed and recognized in machine learning research
(\citealp{GT10}) and bioinformatics society (\citealp{SFK10}). In
protein inference, we will face the same problem as well. The main
objective of this letter is to highlight this fact and people should
be aware of such risk in future comparison when developing new
protein inference algorithms.

\subsubsection*{Funding:}The Natural Science Foundation of China
under Grant No.61003176 and 61073051.
\subsubsection*{Conflict of Interest:}None declared.

\bibliographystyle{bioinformatics}
\bibliography{response}

\end{document}